\documentclass[journal,onecolumn]{IEEEtran}
\usepackage{cite}
\usepackage{setspace}
\ifCLASSINFOpdf
    \usepackage[pdftex]{graphicx}
   \DeclareGraphicsExtensions{.pdf,.jpeg,.png}
\else
   \usepackage[dvips]{graphicx}
   \DeclareGraphicsExtensions{.eps}
\fi
\usepackage[cmex10]{amsmath}
\usepackage{array}
\usepackage[tight,footnotesize]{subfigure}
\usepackage{latexsym, amsfonts, amssymb,wasysym}
\newcommand{\xn}[0]{$\mathbf{x}[n]$ } 
\newcommand{\y}[0]{$\mathbf{y}$ } 
\newcommand{\x}[0]{$\mathbf{x}$ }
\newcommand{\s}[0]{$\mathbf{s}$ }
\newcommand{\psibf}[0]{$\mathbf{\psi}$ } 

\newcommand{\phibf}[0]{$\mathbf{\phi}$ }
 
\DeclareSymbolFont{missing}{OML}{cmr}{m}{n}
\DeclareMathSymbol{\ell}{\mathord}{missing}{'140}

\begin{document}
\title{Compressive Sensing Using the Entropy Functional}

\author{Kivanc Kose, Osman Gunay and A. Enis Cetin}
\maketitle
\begin{abstract}
In most compressive sensing problems $\ell_1$ norm is used during the signal reconstruction process. In this article the use of entropy functional is proposed to approximate the $\ell_1$ norm. A modified version of the entropy functional is continuous, differentiable and convex. Therefore, it is possible to construct globally convergent iterative algorithms using Bregman's row action D-projection method for compressive sensing applications. Simulation examples are presented.
\end{abstract}

\section{Introduction}
Nyquist-Shannon sampling theorem \cite{Shannon} is one of the fundamental theorems in signal processing literature. As it is well known, it specifies the conditions for perfect reconstruction of a continuous signal from its samples. If a signal is sampled with a rate at least two times its bandwidth, it can be perfectly reconstructed from its samples. However in many applications of signal processing including waveform compression, perfect reconstruction is not necessary.

The most common method used in compression applications is the transform coding. The signal \xn is transformed into another domain defined by the transformation matrix \psibf. The transformation procedure is simply finding the inner product of the signal \xn with the rows $\psi_i$ of the transformation matrix \psibf as follows
\begin{equation}
s_i = <\mathbf{x},\psi_i >,\;i = 1,2,...,N,
\end{equation} 
where $\mathbf{x}$ is a column vector, whose entries are samples of the signal \xn. The digital signal \xn can be reconstructed from its transform coefficients $s_i$ as follows;
\begin{equation}
\mathbf{x} = \sum_{i=1}^{N}{s_i . \psi_i} \quad \text{or} \quad \mathbf{x}=\mathbf{\psi}.\mathbf{s}
\label{Trans}
\end{equation}
where $\mathbf{s}$ is a vector containing the transform domain coefficients, $s_i$. The basic idea in digital waveform coding is that the signal should be approximately reconstructed from only a few of its non-zero transform coefficients. In most cases including JPEG image coding standard, the transform matrix \psibf is chosen such that the new signal \s is easily representable in the transform domain with a small number of coefficients. A signal \x is compressible, if it has a few large valued $s_i$ coefficients in the transform domain and the rest of the coefficients are either zeros or very small valued. 

In compressive sensing framework the signal is assumed to be a K-Sparse signal in a transformation domain such as wavelet domain or DCT domain. A signal with length N is K-Sparse, if it has $K$ non-zero and $(N-K)$ zero coefficients in a transform domain. The case of interest in CS problems is when $K<<N$ i.e., sparse in the transform domain. 

The CS theory introduced in \cite{baraniuk, candes, Tao, Donoho} provides answers to the question of reconstructing a signal from its compressed measurements $\mathbf{y}$, which is defined as follows;
\begin{equation}
\mathbf{y} = \mathbf{\phi} . \mathbf{x} = \mathbf{\phi} . \mathbf{\psi} . \mathbf{s} = \mathbf{\theta} . \mathbf{s}
\label{Eq3}
\end{equation}
where \phibf is the $M \times N$ measurement matrix where $M<<N$. The reconstruction of the original signal \x from its compressed measurements \y cannot be achieved by simple matrix inversion or inverse transformation techniques. A sparse solution can be obtained by solving the following optimization problem;
\begin{equation}
\mathbf{s_p} = argmin ||\mathbf{s}||_0 \;\;\; \textit{such that} \;\;\; \mathbf{\theta}.\mathbf{s} = \mathbf{y} . 
\label{CSProblem_l0}
\end{equation}
However this problem is a NP-complete optimization problem therefore its solution can not be found easily. It is also shown in \cite{baraniuk,candes} that, it is possible to construct the \phibf matrix from random numbers which are iid Gaussian random variables and choose the number of measurements as $cKlog(N/K) < M \ll N$ to satisfy the reconstruction conditions defined in \cite{baraniuk} and \cite{candes}. With this choice of the measurement matrix, the optimization problem (\ref{CSProblem_l0}) can be approximated by $\ell_1$ norm minimization as follows,
\begin{equation}
\mathbf{s_p} = argmin ||\mathbf{s}||_1 \;\;\; \textit{such that} \;\;\; \mathbf{\theta}.\mathbf{s} = \mathbf{y}
\label{CSProblem}
\end{equation}

\begin{figure}[!t]
\centering
\includegraphics[width=3.5in]{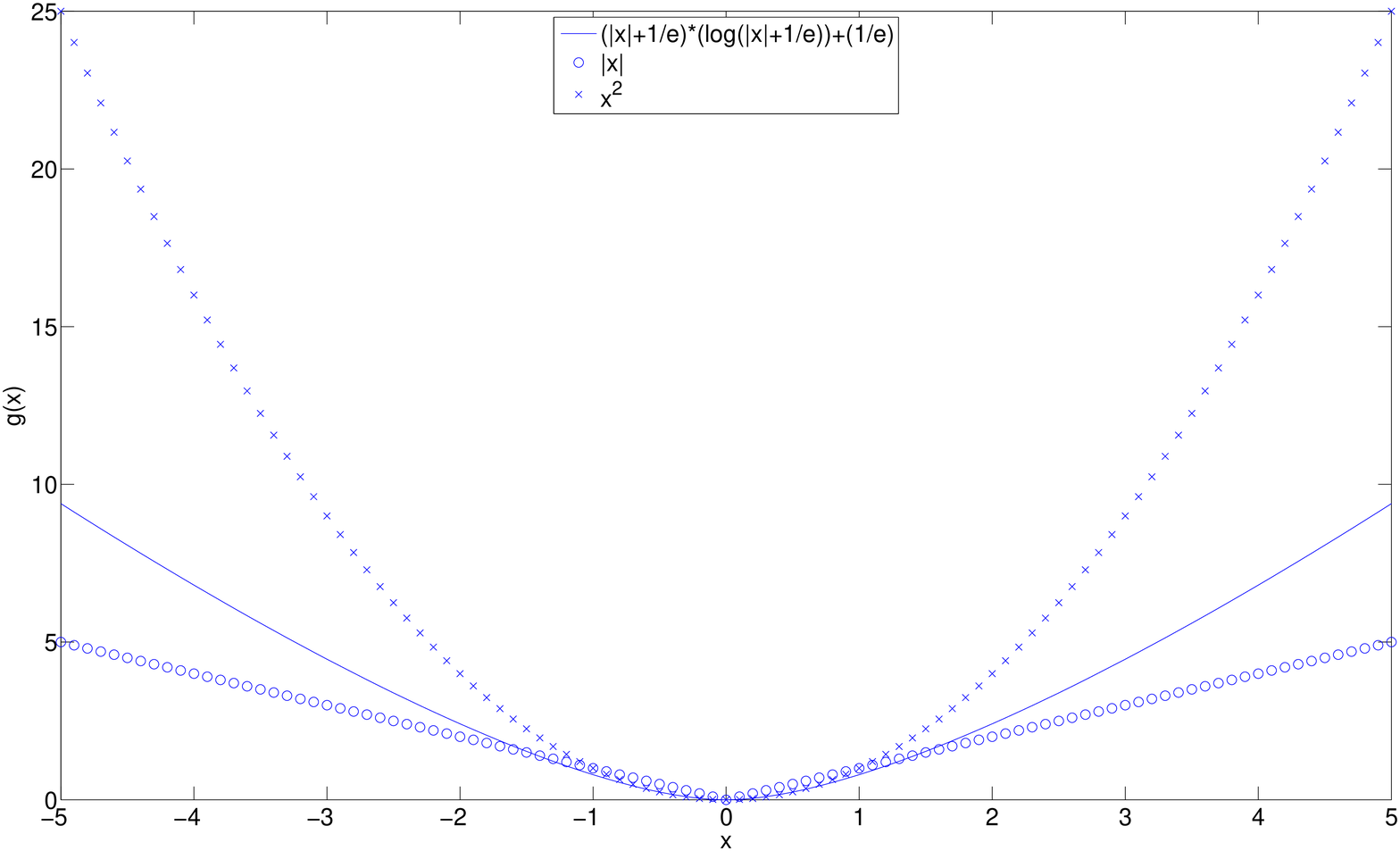}
\caption{Entropy functional $g(v)\;(-)$, $|v| \; (\circ)$ and the Euclidean cost function $v^2 \; (\times)$}
\label{Curves}
\end{figure}

Entropy functional $g(v) = vlogv$ is used to approximately solve some $\ell_1$ optimization problems and linear programming problems in signal and image reconstruction problems by Bregman \cite{Bregman} and others \cite{Herman}, \cite{Enis2, Enis4, Lent} and \cite{Tuy}. 
In this article we propose the use of entropy functional as an alternative way to approximate the CS problem. In Figure \ref{Curves}, plots of the different cost functions including the shifted entropy function
\begin{equation} 
g(v) = (|v|+\frac{1}{e}) \log(|v|+\frac{1}{e})+\frac{1}{e}, 
\label{entropy}
\end{equation}
absolute value $g(v) = |v|$ and $g(v) = v^2$ are shown. The shifted entropy functional (\ref{entropy}) is convex, continuous and differentiable, and it slowly increases compared to $g(v) = v^2$ because log($v$) is much smaller than $v$ for high $v$ values as seen in Figure 1. Bregman also developed iterative row-action methods to solve the global optimization problem by successive local D-Projections. In each iteration step a D-projection, which is a generalized version of the orthogonal projection is performed onto a hyperplane representing a row of the constraint matrix $\mathbf{\theta}$. In \cite{Bregman} Bregman proved that the proposed D-Projection based iterative method is guaranteed to converge to the global minimum regardless of the initial estimate. 

An interesting interpretation of the row-action approach is that it provides a solution to the on-line CS problem. Each new measurement of the signal adds a row to the matrix $\mathbf{\theta}$. In the iterative row-action method a D-projection is performed to the new hyperplane formed by the new measurement. In this way, the currently available solution is updated without solving the entire CS problem. The new solution can be further updated using past measurements or new measurements in an iterative manner by performing other D-Projections. Therefore, it is possible to develop a real-time on-line CS method using the proposed approach. 

The paper is organized as follows. In Section \ref{Algorithm}, we review the D-Projection concept and define the modified entropy functional and related D-Projections. We generalized the entropy function based convex optimization method introduced by Bregman because the ordinary entropy function is defined only for positive real numbers. On the other hand, transform domain coefficients can be both positive and negative. The iterative CS algorithm is explained in Section \ref{iterativealgorithm}. We provide experimental results in Section \ref{Results}.

\section{D-Projection based Algorithm}
\label{Algorithm}

The $\ell_o$ and $\ell_1$ norm based cost functions (\ref{CSProblem_l0}) and (\ref{CSProblem}) used in compressive sensing problems are not differentiable everywhere. In this article we use entropy functional based cost functions to find approximate solutions to the inverse problems defined in
\cite{baraniuk,candes}. Bregman developed convex optimization algorithms in 1960's and his algorithms are widely used in many signal reconstruction and inverse problems \cite{Youla, Herman, Lent, Trussell, Sezan, Combettes, Enis2, Enis3, Enis4, Theodoridis,Cai,Combettes1}.

Assume that the original signal $\mathbf{x}$ can be  represented by a $K$ sparse  length-N vector $\mathbf{s}$ in a transform domain characterized by the transform matrix $\psi$. In CS problems the original signal $\mathbf{x}$ is not available. However $M$ measurements $\mathbf{y}= [y_1, y_2, ..., y_M]^T = \phi \mathbf{x}$ of the original signal is observable via the measurement matrix $\phi$, and the relation between $\mathbf{y}$ and $\mathbf{s}$ are described in Eq. (\ref{Eq3}).

Bregman's method provides  globally convergent iterative algorithms for problems with convex, continuous and differentiable cost functionals $g(.)$:
\begin{equation}
\min_{\mathbf{s}\in C} \; {g(\mathbf{s})}
\label{Dg}
\end{equation}
such that
\begin{equation}
 \theta_i.\mathbf{s} =y_i \quad for \quad i=1,2,...,M.
 \label{hyperplane}
\end{equation}
where $\theta_i$ is the i-th row of the matrix $\theta$. Each equation in (\ref{hyperplane}) represents a hyperplane $H_i$ in $R^N$, which are closed and convex sets in $R^N$. In Bregman's method the iterative reconstruction algorithm starts with an arbitrary initial estimate and successive D-projections are performed onto the hyperplanes $H_i,$ $i=1,2,...,M$  in each step of the iterative algorithm. 

The D-projection onto a closed and convex set is a generalized version of the orthogonal projection onto a convex set \cite{Bregman}. Let  
 $\mathbf{s_o}$ be arbitrary vector in  $R^N$. Its D-Projection $\mathbf{s_p}$ onto a closed convex set $C$ with respect to a cost functional $g(\mathbf{s})$ is defined as follows  
\begin{equation}
\mathbf{s_p} = \arg\;\min_{\mathbf{s}\in C} \; D(\mathbf{s},\mathbf{s_o})
\label{Dproj}
\end{equation}
where
\begin{equation}
D(\mathbf{s},\mathbf{s_o}) =g(\mathbf{s_o})-g(\mathbf{s})-<\bigtriangledown g(\mathbf{s}), \mathbf{s_0}-\mathbf{s})>
\label{Dproj1}
\end{equation}
In CS problems, we have M hyperplanes $H_i: \theta_i.\mathbf{s} =y_i \quad$ for $\quad i=1,2,...,M$.
For each hyperplane $H_i$, the the D-projection (\ref{Dproj}) is equivalent to 
\begin{eqnarray}
\bigtriangledown g(\mathbf{s_p}) = \bigtriangledown g(\mathbf{s_0}) + \lambda \theta_i \\
\theta_i.\mathbf{s_p} = y_i
\label{lagrangian}
\end{eqnarray}
where $\lambda$ is the Lagrange multiplier. As pointed above
the D-projection is a generalization of the orthogonal projection. When the cost functional is the Euclidean cost functional $g(\mathbf{s})= \sum_n s(n)^2$ the distance $D(\mathbf{s_1}, \mathbf{s_2})$ becomes the $\ell_2$ norm of difference vector $(\mathbf{s_1}-\mathbf{s_2})$, and the D-projection simply becomes the well-known orthogonal projection onto a hyperplane.

The orthogonal projection of an arbitrary vector 
$\mathbf{s_o}=[s_0[1],s_0[2],...,s_0[M]]$ onto the hyperplane $H_i$ is given by
\begin{equation}
s_p(n) = s_0(n)+\lambda \theta_i(n) , n = 1,2,...,N 
\label{s_p}
\end{equation}
where $\theta_i(n)$ is the n-th entry of the vector  $\theta_i$ and the Lagrange multiplier $\lambda$ \; is given by,
\begin{equation}
\lambda = {{y_i-\sum^{N}_{n=1}s_0(n) \theta(i,n)}\over{\sum^{N}_{n=1}{\theta_i}{^2}(n)}}
\label{lambda}
\end{equation}
When the cost functional is the 
entropy functional $g(s)= \sum_n s(n) \log(s(n))$, the D-projection onto the hyperplane $H_i$ leads to the following equations
\begin{equation}
s_p(n) = s_o(n).e^{(\lambda .\theta_i(n))},\; n = 1,2,...,N
\label{entropicupdate1}
\end{equation}
where the Lagrange multiplier $\lambda$ is obtained by inserting (\ref{entropicupdate1}) into the following hyperplane equation:
\begin{equation}
\theta_i \mathbf{s} = y_i
\end{equation}
because the D-projection $\mathbf{s_p}$ must be on the hyperplane $H_i$. The above set of equations are used in signal reconstruction from Fourier Transform samples \cite{Enis4} and the tomographic reconstruction problem \cite{Herman}. The entropy functional is defined only for positive real numbers. In CS problems entries of vector $\mathbf{s}$ can take both positive and negative values. We modify the entropy functional and extend it to negative real numbers as follows:
\begin{equation}
\min \sum^{N}_{i=1}(|s_i|+\frac{1}{e}).(log(|s_i|+\frac{1}{e})) \quad s.t. \quad \theta . \mathbf{s} = y.
\label{entoroptim}
\end{equation}
where subscript $e$ represents the term entropy. The continuous cost functional $g_e(s)$ satisfies the following conditions,

(i) $\frac{\partial g_e}{\partial s_i}(0) = 0$, $i = 1,2,...,N$ and

(ii) $g_e$ is strictly convex and continuously differentiable.

On the other hand, the $\ell_1$ norm is unfortunately not a globally smooth function therefore it can be solved using non differentiable minimization techniques such as sub-gradient methods \cite{Hirriart}. Another way of approximating the $\ell_1$ penalty function using an entropic functional is available in \cite{MPinar}. 

To obtain the D-projection of $\mathbf{s_o}$ onto a hyperplane $H_i$ with respect to  the entropic cost functional (\ref{entoroptim}), we need to minimize the generalized distance $D(\mathbf{s},\mathbf{s_o})$ between $\mathbf{s_0}$ and the hyperplane $H_i$:
\begin{equation}
D(\mathbf{s},\mathbf{s_o}) =  \mathbf{g_e(s_o)}- \mathbf{g_e(s)}+< \bigtriangledown \mathbf{g_e(s)}, \mathbf{s}-\mathbf{s_o}>
\label{entropicupdate}
\end{equation}
with the condition that $\theta_i \mathbf{s} = y_i$. Using (\ref{lagrangian}), the values of $s$ can be obtained as;

\begin{equation}
s(n) = (s_o(n)+sgn(s(n)).{e^{-1}}).e^{(sgn(s(n)).\lambda .\theta_i(n))}-(sgn(s(n)).e^{-1}), \quad n=1,2,...,N.
\label{entropicLagrange}
\end{equation}
where $\lambda$ is the Lagrange multiplier.

The D-projection vector and the $\mathbf{s_*}$ satisfies the set of equations (\ref{entropicLagrange}), and the hyperplane equation $H_i: \theta_i.\mathbf{s} = y_i$.

\subsection{Iterative Reconstruction Algorithm}
\label{iterativealgorithm}
The global convex optimization problem defined in (\ref{entoroptim}) is solved by performing successive local D-projections onto hyperplanes defined by the rows of the matrix $\theta$. 

\begin{figure}[!t]
\centering
\includegraphics[width=3 in]{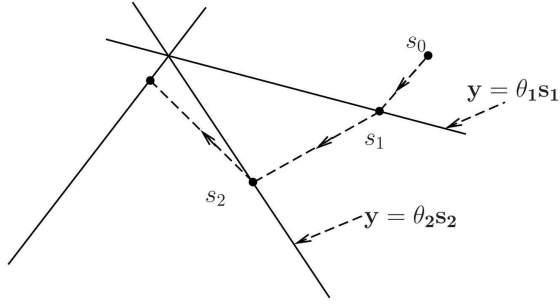}
\caption{Geometric interpretation of the entropic projection method: Sparse representation $\mathbf{s_i}$ corresponding to decision functions at each iteration are updated so as to satisfy the hyperplane equations defined by the measurements $y_i$ and the measurement vector $\theta_i$. Lines in the figure represent hyperplanes in $\mathbb{R}^N$. Sparse representation vector $s_i$ converges to the intersection of the hyperplanes. Notice that D-projections are not orthogonal projections.}
\label{ProjectionFig}
\end{figure}

The iterations start with an arbitrary initial estimate  $\mathbf{s_o}$. This vector is D-projected onto the hyperplane $H_1$ and $\mathbf{s_1}$ is obtained. The iterate $\mathbf{s_1}$ is projected onto the next hyperplane $H_2$ (see Figure \ref{ProjectionFig})... ${N-1}^{st}$ estimate $\mathbf{s}_{N-1}$ is D-projected onto $H_N$ and $\mathbf{s}_{N}$ is obtained. In this way the first iteration cycle is completed. The vector $\mathbf{s}_{N}$ is then projected onto the hyperplane $H_1$ and $\mathbf{s}_{N+1}$ is obtained etc. Bregman proved that $s_i$ defined in (\ref{entoroptim}) converges to the solution of the optimization problem.

\section{Experimental Results}
\label{Results}
\begin{figure}[!t]
\centering
\includegraphics[width=3.5in]{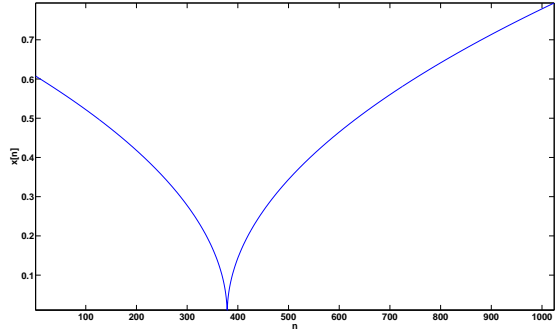}
\caption{The cusp signal with $N=1024$ samples}
\label{Cusp}
\end{figure}

\begin{figure}[!t]
\centering
\includegraphics[width=3.5in]{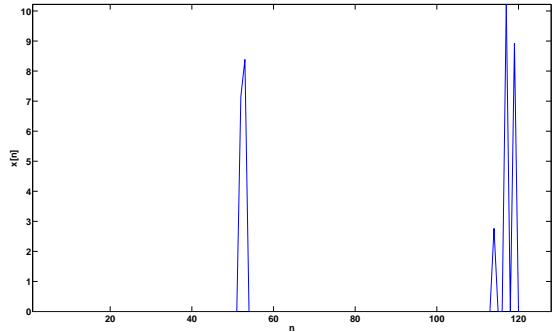}
\caption{Random S=4 sparse signal with $N=128$ samples}
\label{RS}
\end{figure}

For the validation and testing of the entropic minimization method, two experiments with two different signals are carried out. The \textit{cusp} signal (Figure \ref{Cusp}), which has 1024 samples, and the \textit{random sparse} (Figure \ref{RS}) signal, which has 128 samples, are used. The cusp signal is $S = 72$ sparse in DCT domain and the random signal has four non-zero samples.

\begin{figure*}[!t]
\centering
\subfigure[$N=1024$ length cusp signal reconstructed from $2S=144$ measurements]{
\includegraphics[width=7in]{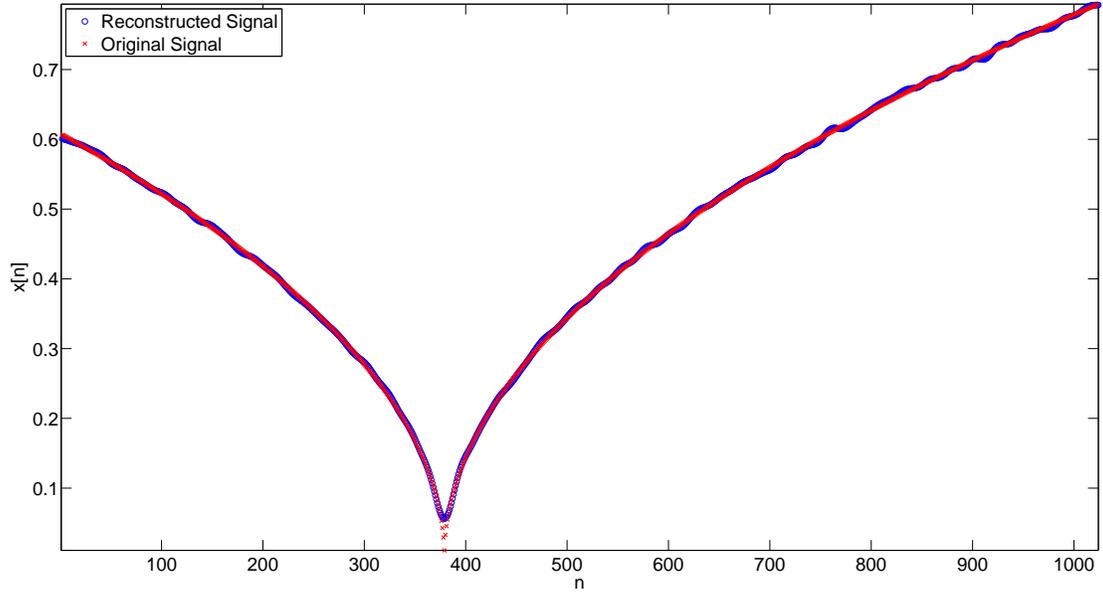}
\label{CuspRec2}}
\subfigure[$N=1024$ length cusp signal reconstructed from $10S=720$ measurements]{
\includegraphics[width=7in]{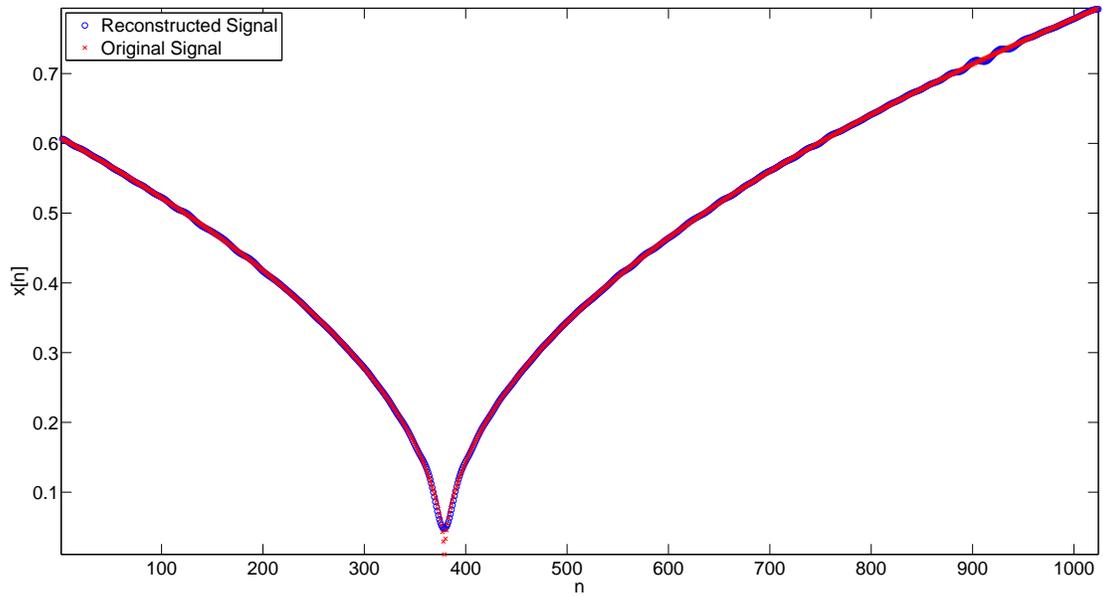}
\label{CuspRec10}}
\caption{The cusp signal with 1024 samples reconstructed from $M = 2S$ (a) and $M =10S$ (b) measurements using the iterative, entropy functional based method.}
\end{figure*}

\begin{figure*}[!t]
\centering
\subfigure[$N=128$ length random sparse signal reconstructed from $6S=24$ measurements]{
\includegraphics[width=7in]{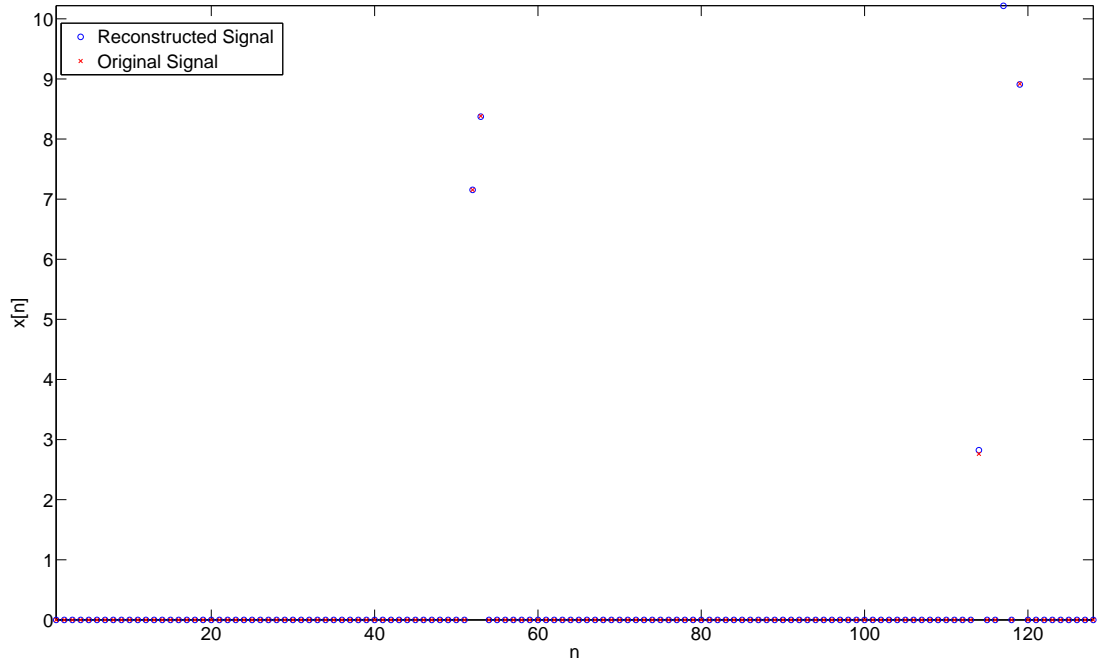}
\label{RandRec6}}
\subfigure[$N=128$ length random sparse signal reconstructed from $10S=40$ measurements]{
\includegraphics[width=7in]{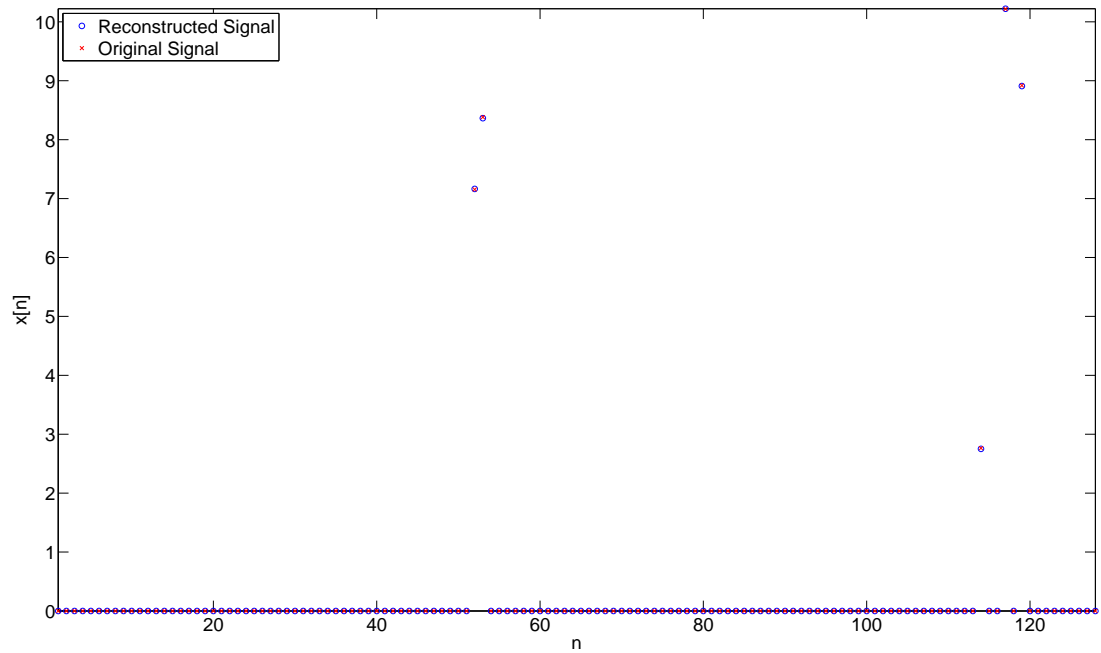}
\label{RandRec10}}
\caption{Random sparse signal with 128 samples is reconstructed from (a) $M = 6S$ and (b) $M =10S$ measurements using the iterative, entropy functional based method.}
\end{figure*}

\begin{figure*}[!t]
\centering
\subfigure[The DCT of the cusp signal reconstructed from its measurements using pseudo-inversion v.s. the DCT of the original cusp signal.]{
\includegraphics[width=7in]{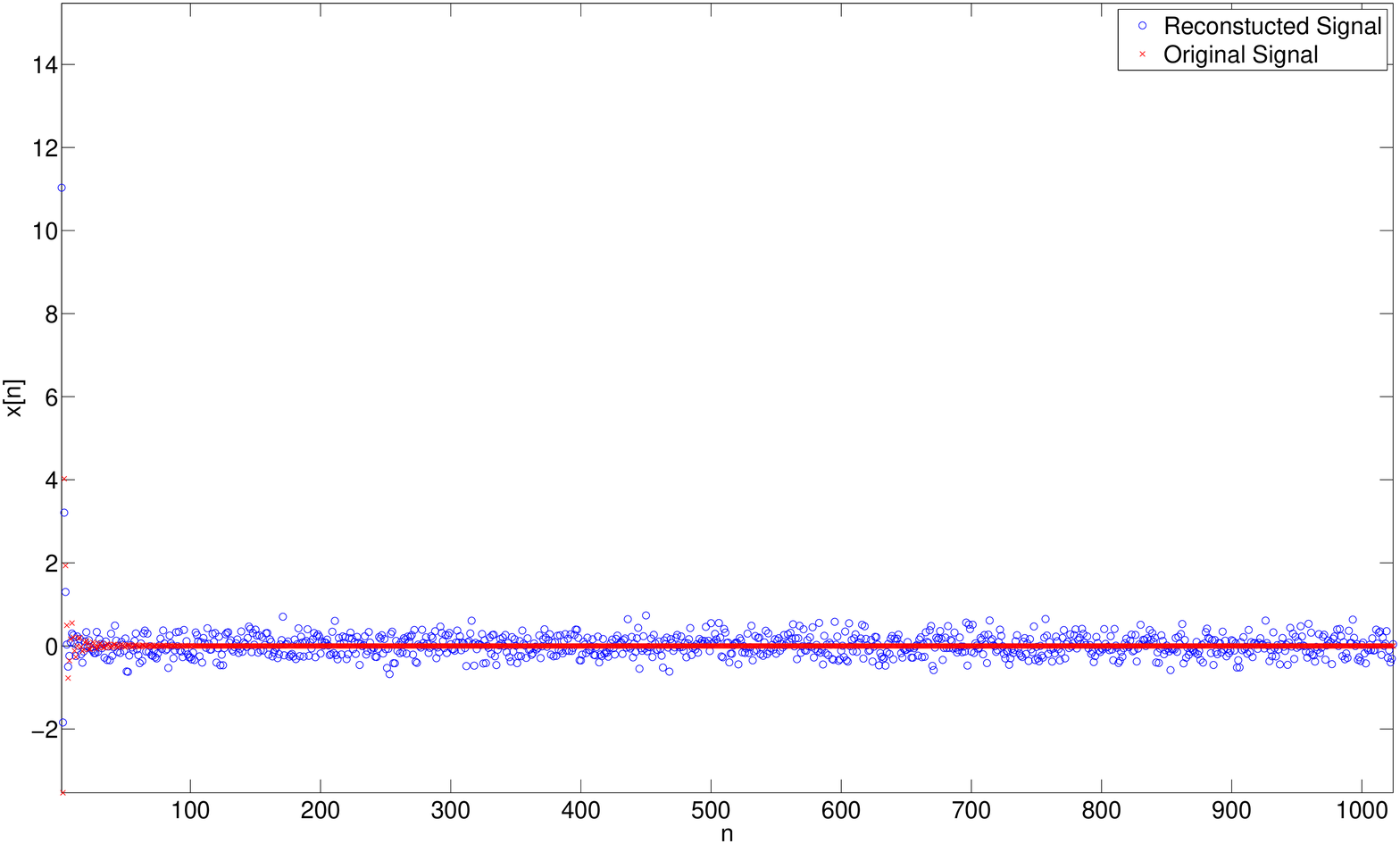}
\label{invCusp}}
\subfigure[Random sparse signal reconstructed from its measurements using pseudo-inversion v.s. the original signal.]{
\includegraphics[width=7in]{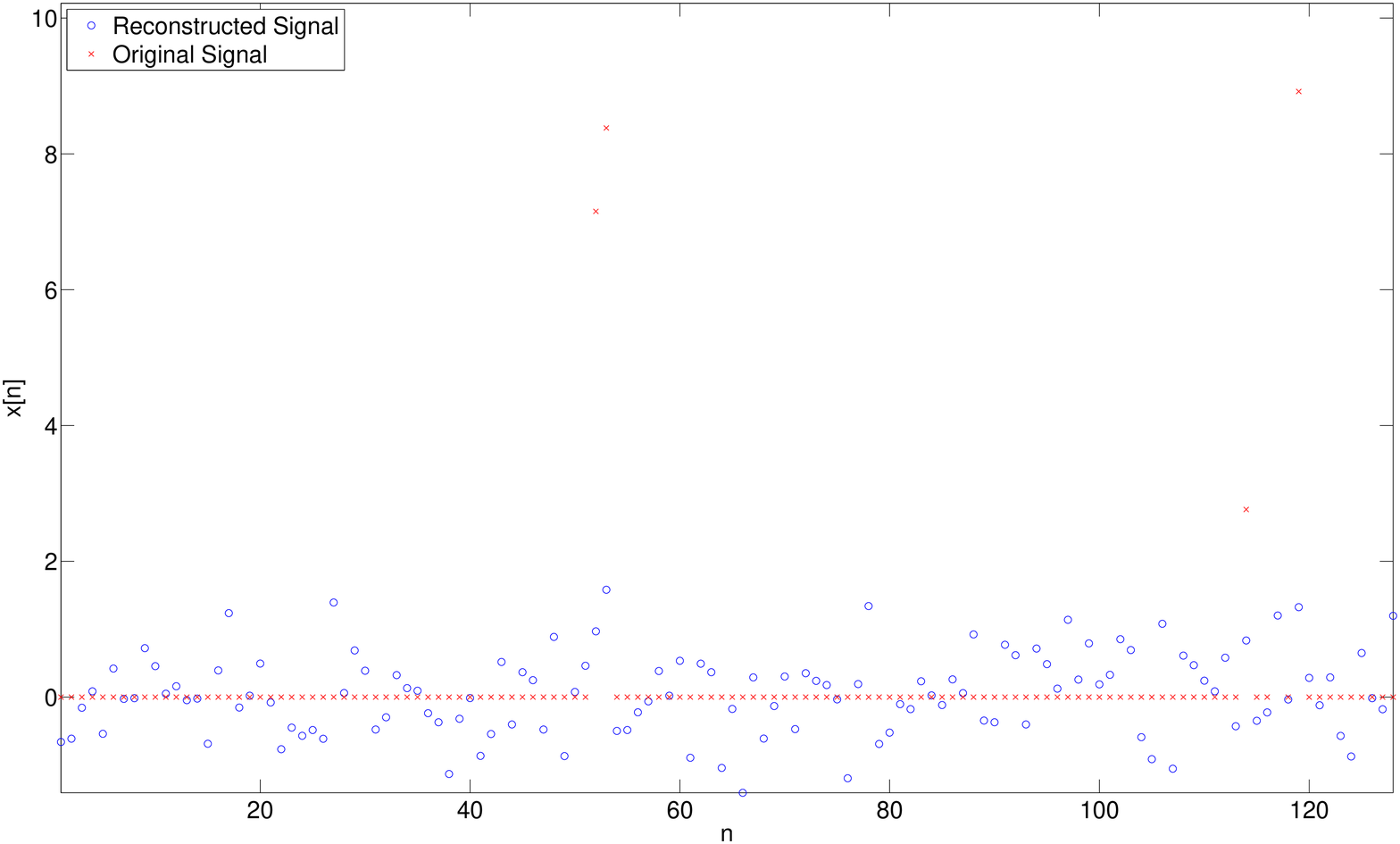}
\label{invRand}}
\caption{The DCT of the cusp signal and the random sparse signal reconstructed from their measurements using pseudo-inversion.}
\end{figure*}

The measurement matrices $\phi$ are chosen as Gaussian random matrices. In the experiments $M = 2S$ and $M =10S$  measurements are taken from the cusp signal and $M = 6S$ and $M =10S$  measurements are taken from the random signal. The original signals are reconstructed from those measurements. The reconstructed signals using the iterative method using entropy based cost functional are shown in Figures \ref{CuspRec2},  \ref{CuspRec10}, \ref{RandRec6}, and \ref{RandRec10}. The reconstructed signals using the pseudo-inverse of the $\theta$ matrix are shown in Figures \ref{invCusp}, and \ref{invRand} respectively.

\section{Conclusion and Future Work}
In this article we present; (i) the entropy based cost function for solving the CS problems, and (ii) an iterative row-action method to solve the CS problem. The entropy based cost functional estimates the $\ell_1$ norm. Both the theoretical results given in Section \ref{Algorithm} and the experimental results given in Section \ref{Results} indicate that the entropy based cost function and the iterative row-action method can be used for reconstructing a sparse signal from its measurements. 


It is also shown that the row action methods provide a solution to the on-line CS problem. The reconstruction result can be updated on-line according to the new measurements without solving the entire optimization problem again in real time.

We will compare this entropy functional approach with other methods available including the TV approach \cite{Combettes2,Herman2}


\clearpage

\begin{thebibliography}{99}

\bibitem{Shannon}
C. E. Shannon, ``Communication in the presence of noise", \textit{Proceedings of Institute of Radio Engineers}, vol. 37, no. 1, pp. 10-21, Jan. 1949.
\bibitem{baraniuk} 
R. G. Baraniuk, ``Compressed sensing [Lecture Notes]", \textit{IEEE Signal Processing Magazine}, vol 24, no: 4, pp.118-124, July, 2007
\bibitem{candes} 
E. Candes, J. Romberg, T. Tao, ``Robust uncertainty principles: exact signal reconstruction from highly incomplete frequency information", \textit{IEEE Transactions on Information Theory}, vol.52, no.2, pp. 489- 509, February 2006
\bibitem{Tao} 
E. Candes and T. Tao, ``Near optimal signal recovery from random projections: Universal encoding strategies?", \textit{IEEE Trans. Inform. Theory}, vol. 52, no. 12, pp. 5406–5425, Dec. 2006.
\bibitem{Donoho} 
D. Donoho, ``Compressed sensing", \textit{IEEE Transactions on Information Theory}, vol. 52, no. 4, pp. 1289-1306, Apr. 2006.
\bibitem{Bregman} 
L.M.Bregman. ``The Relaxation method of finding the common point of convex sets and its applications to the solution of the problems in convex programming", \textit{USSR Computational Mathematics, Mathematical Physics}, vol 7 pp 200-217, 1967
\bibitem{Hirriart} 
J-B., Hirriart-Urruty, and C. Lemarachal, ``Convex Analysis and Minimization Algorithms II", \textit{Springer Verlag}, Berlin, 1993.
\bibitem{MPinar} 
M. C., Pinar,  S. A. Zenios, ``An Entropic Approximation of $\ell_1$ Penalty Function", \textit{Transactions on Operational Research}, vol. 7, pp.101-120, 1995
\bibitem{Youla}
D. C. Youla, H. Webb, ``Image restoration by the method of convex projections: part 1 theory", \textit{IEEE Transactions on Medical Imaging}, Vol. 1 No. 2 pp. 81-94, 1982.
\bibitem{Trussell} 
H . Trussell, M. Civanlar, ``The Landweber iteration and projection onto convex sets," \textit{IEEE Transactions on Acoustics, Speech and Signal Processing}, vol.33, no.6, pp. 1632- 1634, Dec 1985
\bibitem{Sezan} 
M. I. Sezan, H. Stark, ``Image Restoration by the Method of Convex Projections: Part 2-Applications and Numerical Results," \textit{IEEE Transactions on Medical Imaging}, vol.1, no.2, pp.95-101, Oct. 1982
\bibitem{Combettes} 
P. L. Combettes, ``The foundations of set theoretic estimation", \textit{Proceedings of the IEEE}, vol. 81, No.2, pp.182-208, Feb. 1993
\bibitem{Herman} 
G. T., Herman, ``Image Reconstruction From Projections", \textit{Real-Time Imaging}, vol. 1, no.1, pp. 3-18, 1995
\bibitem{Lent} 
Y. Censor and A. Lent, ``An iterative row-action method for interval convex programming", \textit{Journal of Optimization theory and Applications}, vol. 34, no. 3, pp. 321-353, 1981, DOI: 10.1007/BF00934676
\bibitem{Theodoridis} 
S. Theodoridis, K. Slavakis, I. Yamada, ``Adaptive Learning in a World of Projections", \textit{IEEE Signal Processing Magazine}, vol.28, no.1, pp.97-123, Jan. 2011, DOI: 10.1109/MSP.2010.938752
\bibitem{Tuy} 
A. Lent, H. Tuy, ``An iterative method for the extrapolation of band-limited functions",  \textit{Journal of Mathematical Analysis and Applications}, 83 (2), pp. 554-565, 1981
\bibitem{cetin} 
A. E. Cetin, ``Reconstruction of signals from Fourier transform samples," \textit{Signal Processing}, vol. 16, pp. 129–148, 1989.
\bibitem{Enis2}
A. E. Cetin, `An Iterative Algorithm for Signal reconstruction from Bispectrum,", \textit{IEEE Transactions Signal Processing}, vol. 39, no. 12, pp. 2621-2628, Dec. 1991
\bibitem{Enis3}
A. E. Cetin, R. Ansari, ``Signal recovery from wavelet transform maxima", \textit{IEEE Transactions on Signal Processing}, vol. 42, No.1, pp. 194-196, 1994.
\bibitem{Enis4} 
A. E. Cetin and R. Ansari, ``Convolution-based framework for signal recovery and applications", \textit{Journal of Optical Society of America}, A vol. 5, Iss. 8, pp. 1193-1200,1988.
\bibitem{Cai} J. F. Cai, S. Osher,Z. Shen, ``Linearized Bregman iterations for compressed sensing", \textit{Mathematics and Computation}, 78, 1515-1536 2009. 
\bibitem{Combettes1} P. L. Combettes and J. C. Pesquet, ``Proximal thresholding algorithm for minimization over orthonormal bases", \textit{SIAM Journal of Optimization}, 18, pp. 1351–1376, 2007
\bibitem{Combettes2} P. L. Combettes, J. C. Pesquet, ``Image restoration subject to a total variation constraint". \textit{IEEE Transactions on Image Processing} 13, 1213–1222, 2004.
\bibitem{Herman2} R. Davidi, G. Herman, and Y. Censor, ``Perturbation-resilient block-iterative projection methods with application to image reconstruction from projections". \textit{International Transactions in Operational Research}, 16: 505–524, 2009.
\end{thebibliography}
\end{document}